\documentclass[letterpaper]{article} 

\usepackage{aaai24}  
\usepackage{times}  
\usepackage{helvet}  
\usepackage{courier}  
\usepackage[hyphens]{url}  
\usepackage{graphicx} 
\usepackage{natbib}  
\usepackage{caption} 
\usepackage{amsthm,amsmath,amssymb,amsfonts}
\usepackage{booktabs}

\title{SAT-Based Techniques for Lexicographically Smallest Finite Models}
\newcommand{\miko}{Mikol\'a\v{s} Janota}
\newcommand{\mike}{Michael Codish}
\newcommand{\joao}{Jo\~ao Ara\'ujo}
\newcommand{\ccw}{Choiwah Chow}
\newcommand{\petr}{Petr Vojt\v{e}chovsk\'y}

\author {
  \miko\textsuperscript{\rm 1}, \ccw\textsuperscript{\rm 1}, \joao\textsuperscript{\rm 2, 3}, \mike\textsuperscript{\rm 4}, \petr\textsuperscript{\rm 5}
}

\affiliations {
    \textsuperscript{\rm 1}Czech Technical University in Prague, Czechia\\
    \textsuperscript{\rm 2}Center for Mathematics and Applications (NOVA Math), Portugal\\
    \textsuperscript{\rm 3}Department of Mathematics, NOVA FCT, Portugal\\
    \textsuperscript{\rm 4}Ben-Gurion University of the Negev, Beer-Sheva, Israel\\
    \textsuperscript{\rm 5}University of Denver, USA\\
    mikolas.janota@cvut.cz
}

\theoremstyle{plain}

\newtheorem{theorem}{Theorem}
\newtheorem{proposition}[theorem]{Proposition}
\theoremstyle{definition}
\newtheorem{definition}[theorem]{Definition}
\newtheorem{observation}[theorem]{Observation}
\newtheorem{example}[theorem]{Example}

\usepackage[ruled,vlined]{algorithm2e}
\DontPrintSemicolon%
\SetKwFunction{SAT}{SAT}
\SetKwFunction{encode}{enc}
\SetKwFunction{bij}{X}
\newcommand{\xpi}[2]{x_{{#1}\rightarrow{#2}}} 
\newcommand{\xn}[3]{x_{{#1},{#2}:{#3}}} 
\newcommand{\SRC}{\ast}
\newcommand{\lexmin}{\textsc{LexMin}\xspace}

\newcommand{\DST}{\diamond}
\newcommand{\src}[2]{{#1}\SRC{#2}}
\newcommand{\dst}[2]{{#1}\DST{#2}}
\newcommand{\down}{\downarrow\!\!}
\usepackage{xspace}

\newcommand{\nauty}{\texttt{nauty}\xspace}
\newcommand{\mlex}{\texttt{mlex}\xspace}

\usepackage{tikz}
\usetikzlibrary{matrix}
\usetikzlibrary{calc,patterns}
\tikzstyle{finmod} = [%
		font={},
		column 1/.style={font=\bfseries},
		row 1/.style={font=\bfseries},
		inner sep=0pt, 
    matrix of nodes,draw=none,nodes={%
    text width=1.2em,align=center, minimum height=1.0em,anchor=center}
]

\begin{document}
\maketitle

\begin{abstract}
This paper proposes SAT-based techniques to calculate a specific normal form of a given finite mathematical structure (model). The normal form is obtained by permuting the domain elements so that the representation of the structure is lexicographically smallest possible. Such a normal form is of interest to mathematicians as it enables easy cataloging of algebraic structures. In particular, two structures are isomorphic precisely when their normal forms are the same. This form is also natural to inspect as mathematicians have been using it routinely for many decades.

We develop a novel approach where a SAT solver is used in a black-box fashion to compute the smallest representative. The approach constructs the representative gradually and searches the space of possible isomorphisms, requiring a small number of variables. However, the approach may lead to a large number of SAT calls and therefore we devise propagation techniques to reduce this number. The paper focuses on finite structures with a single binary operation (encompassing groups, semigroups, etc.). However, the approach is generalizable to arbitrary finite structures. We provide an implementation of the proposed algorithm and evaluate it on a variety of algebraic structures.
\end{abstract}

\section{Introduction}
Finite model finding of first-order or higher-order logic has a long-standing
tradition in automated reasoning. A number of techniques
have been researched in
SAT~\cite{claessen03}, constraint
programming~\cite{AudemardBH06,ZhangJ1996Falcon,ZhangZ95},
or SMT~\cite{reynolds-cav13,reynolds-cade13}.
In theorem proving and software verification, finite models are typically used
to identify incorrectly stated theorems. In computational algebra,
mathematicians use finite model finding to study fundamental algebraic
structures.

This paper does not focus on calculating specific models but
on providing a normal form for a given model.
This is one of the most prevalent problems in mathematics, i.e., assigning a
canonical representative to an equivalence class. For example, the canonical
form of a rational fraction is the quotient with the common prime factors
removed (reduced fraction); Jordan's canonical form for matrices assigns a
matrix to an equivalence class of similar matrices; there are ways of assigning
a canonical form to a graph so that any two are isomorphic if and only if their
canonical forms are the same, etc. But helping with decision problems is just
one of the applications of canonical forms.  When we want to
enumerate all structures of a given type (e.g., all triangulated 3-manifolds)
up to some size (e.g., on 11 vertices~\cite{Lutz2008,lutz2009isomorphismfree}),
it suffices to generate the canonical forms and ignore all the rest. These
are just a few examples as the applications of canonical forms are countless,
including applications to topics as far as
chemistry~\cite{Weininger1989,Schneider2015}. The key feature of canonical
systems of representatives is that two objects belong to the same equivalence
class if and only if their canonical forms are equal.

A largely widespread technique to assign canonical forms to mathematical
objects is to associate each object in the class with a vector and then order
all the vectors lexicographically: the canonical object will be the object with
the smallest vector. We will call this the \emph{lexicographically
smallest representative}, \emph{lexmin} for short. In constraint programming
literature, a related term \emph{lex-leader} is
defined, cf.~\citet{walsh-aaai12,cp-handbook}. Lexmin for graphs is also
extensively studied in the literature, cf.~\citet{babai83,crawford-kr96}.

In computational algebra, this idea naturally translates to concatenating the
rows of a multiplication table into a single vector. This canonical
form was used as early as 1955 to calculate all the distinct\footnote{Two
  semigroups are distinct if they cannot be mapped to one another by an
isomorphism or by an anti-isomorphism.}
semigroups of order 4. More recently, Jipsen maintains an online
database of a variety of mathematical structures stored as
lexmin~\cite{structures}, the GAP package Smallsemi enables calculating lexmin semigroups~\cite{smallsemi}.



\begin{figure}[t]
  \centering
  \begin{tikzpicture}
    \matrix [finmod] (SRC) at (0, 0) {
      $\SRC$   & 1 & 2 & 3 & 4 & 5 & 6 & 7 \\
      1  & 7 & 5 & 6 & 1 & 4 & 2 & 3 \\
      2  & 5 & 3 & 1 & 2 & 6 & 7 & 4 \\
      3  & 6 & 1 & 5 & 3 & 7 & 4 & 2 \\
      4  & 1 & 2 & 3 & 4 & 5 & 6 & 7 \\
      5  & 4 & 6 & 7 & 5 & 2 & 3 & 1 \\
      6  & 2 & 7 & 4 & 6 & 3 & 1 & 5 \\
      7  & 3 & 4 & 2 & 7 & 1 & 5 & 6 \\
    };

    \matrix [finmod] (DST) at (4.3, 0) {
      $\DST$ & 1 & 2 & 3 & 4 & 5 & 6 & 7 \\
      1  & 1 & 2 & 3 & 4 & 5 & 6 & 7 \\
      2  & 2 & 3 & 4 & 5 & 6 & 7 & 1 \\
      3  & 3 & 4 & 5 & 6 & 7 & 1 & 2 \\
      4  & 4 & 5 & 6 & 7 & 1 & 2 & 3 \\
      5  & 5 & 6 & 7 & 1 & 2 & 3 & 4 \\
      6  & 6 & 7 & 1 & 2 & 3 & 4 & 5 \\
      7  & 7 & 1 & 2 & 3 & 4 & 5 & 6 \\
    };

    \foreach \name in {SRC,DST} {%
      \draw[gray] (\name-8-1.south east) to (\name-1-1.north east); 
      \draw[gray] (\name-1-1.south west) to (\name-1-8.south east); 
    }
  \end{tikzpicture}
  \caption{$(D, \SRC)$ and its lexmin $(D,\DST)$ for $D=\{1..7\}$.}\label{fig:example}
\end{figure}

Figure~\ref{fig:example} shows a motivating example of a possible
multiplication table for an operation $\SRC$ together with its
lexicographically smallest representative $\DST$. It is relatively easy for
a human to detect that $\SRC$ is a quasigroup (aka Latin square), however,
further properties are harder to see. In contrast, the multiplication table of
$\DST$ is much easier to comprehend---we see that the operation corresponds to
addition modulo 7, which is in fact the unique group of order~7 (the cyclic
group $\mathbb{Z}_7$).

Developing efficient algorithms for calculating the lexmin form is
paramount in the field of computational algebra:

\begin{itemize}
  \item
    It enables presenting a concrete algebra in a familiar way to researchers.
  \item
    Computational algebra systems, such as GAP~\cite{GAP4}, contain a large
    number of packages for handling algebras for specific forms and lexmin
    provides a uniform exchange format between these packages.
  \item
    Lexmin provides a uniform way of storing and recalling algebras. The form
    is especially interesting for prefix trees (tries) since inherently, many
    algebras will share the same prefix in the lexmin form.
\end{itemize}

\noindent This paper presents the following contributions.
\begin{itemize}

  \item We develop a SAT-based algorithm that enables calculating the normal
    form on the fly, rather than working with explicit representation of the
    target normal form.

  \item We design a variety of propagation techniques
    that enable avoiding SAT calls in a large number of cases, which has
    proven indispensable in many real-world problems.

  \item We provide a prototype implementation of the proposed algorithm, using
    state-of-the-art SAT solvers in a black box fashion. This prototype is
    evaluated on a number of algebras that mathematicians deal with on daily
    basis.

\end{itemize}

\section{Preliminaries}\label{sec:preliminaries}

Throughout the paper we focus on finite mathematical structures with a single
binary operation, hereafter referred to as \emph{magmas} (the term
\emph{groupoid} is also used in the literature). For instance, any finite group
or semigroup is a magma. Magmas are denoted by a pair $(D,\circ)$ where $D$ is
the domain and $\circ$ a binary operation on $D$. We rely on the
well-established term of isomorphism.

\begin{definition}[isomorphism]\label{def:isomorphism}
A bijection $f:D_1\rightarrow D_2$ is an isomorphism from a magma $(D_1,\SRC)$ to
$(D_2,\DST)$ if $f(\src{a}{b})=\dst{f(a)}{f(b)}$, for all $a,b\in D_1$. Two
magmas are \emph{isomorphic} iff there exists at least one isomorphism between them.
\end{definition}

Throughout the paper, we consider a finite domain
$D=\{1,\ldots,n\}$ for $n\in\mathbb{N}^+$. The goal is to obtain the
lexicographically smallest $(D,\DST)$ isomorphic to the given 
$(D,\SRC)$.

\begin{definition}[$\preceq$]
  Define a total order $\preceq$ on magmas on domain $D$ as follows.
  For magmas $A=(D,\SRC)$ and $B=(D, \DST)$, we have $A\preceq B$ 
  iff 
  $\src{1}{1},\src{1}{2},\dots,\src{1}{n},\src{2}{1},\dots,\src{n}{n}$
  is lexicographically smaller or equal to
  $\dst{1}{1},\dst{1}{2},\dots,\dst{1}{n},\dst{2}{1},\dots,\dst{n}{n}$.
\end{definition}

\begin{definition}[\lexmin]
  For magma $A=(D,\SRC)$, magma $B=(D, \DST)$ is the \emph{lexicographically
  smallest representative (lexmin)} of $A$ iff $B$ is the $\preceq$-least element among
  all magmas $(D,\DST')$ isomorphic to $A$.
  The $\lexmin$ problem is finding the lexicographically smallest representative of $A$.
\end{definition}

\noindent In several cases we rely on the notion of an idempotent, which is invariant under isomorphism.

\begin{definition}[idempotent]
  For a magma $(D,\SRC)$, an element $a\in D$ is an \emph{idempotent} iff $\src{a}{a}=a$.
\end{definition}

\begin{observation}
  Let $A=(D_1,\SRC)$ and $B=(D_2,\DST)$ be isomorphic magmas under some
  isomorphism $f$, and let $a$ be an idempotent of $A$, then $f(a)$ is
  an idempotent of $B$.
\end{observation}

\begin{example}\label{ex:iso}
  This example shows a multiplication table for a small magma $(D,\SRC)$ with $D=\{1,2\}$
  together with an extensive representation as a set of assignments.
  On the right-hand side, we see its lexicographically smallest representative~$\DST$.
  The corresponding isomorphism swaps 1 and 2, i.e., $f(1)=2, f(2)=1$,
  alternatively represented as a permutation in the cyclic notation~$(1\,2)$. 

\begin{center}
\begin{tikzpicture}
	\matrix [finmod] (B) at (0, 0) {
    $\mathbf{\SRC}$ & 1 & 2\\
                      1 & 1 & 2\\
                      2 & 2 & 2\\
	};

  \matrix[matrix of nodes,inner sep=2pt] (BL) at (1.75, 0) {
    1&$\SRC$&1&$=$&1\\
    1&$\SRC$&2&$=$&2\\
    2&$\SRC$&1&$=$&2\\
    2&$\SRC$&2&$=$&2\\
	};
  \matrix[matrix of nodes,inner sep=2pt] (AL) at (5.75, 0) {
    2&$\DST$&2&$=$&2\\
    2&$\DST$&1&$=$&1\\
    1&$\DST$&2&$=$&1\\
    1&$\DST$&1&$=$&1\\
	};

	\matrix [finmod] (A) at (4, 0) {
    $\mathbf{\DST}$ & 1 & 2\\
                      1 & 1 & 1\\
                      2 & 1 & 2\\
	};

	\foreach \name in {A,B} {%
		\draw[gray] (\name-3-1.south east) to (\name-1-1.north east); 
		\draw[gray] (\name-1-1.south west) to (\name-1-3.south east); 
	}
\end{tikzpicture}
\end{center}

   Note that the isomorphism not only changes the contents of the
   table but also permutes rows and columns. In this example, to
   obtain $\DST$ from $\SRC$ means swapping rows 1 and 2, columns 1 and 2, 
   and values 1 and 2 in the table.
\end{example}

Example~\ref{ex:iso} also illustrates that properties based on equality are
preserved: both tables contain a row with all elements distinct, have 2
idempotents, etc. This is a more general property, which we state here informally.\footnote{%
  More precisely, a set $S$
  defined by an FOL formula in a magma $A$ corresponds to the set $f(A)$ in $B$ for an
  isomorphism $f$ from $A$ to $B$, \textit{cf.}~Theorem~1.1.10
  in~\cite{Marker2003-MARMTA-14}.
}

\begin{observation}\label{observation:invariant}
  Any property of $A=(D,\DST)$ that does not rely on the names of elements of $D$
  is preserved in all isomorphic copies of $A$.
\end{observation}

Note that in the small Example~\ref{ex:iso}, there is a unique isomorphism from
the input magma to its lexmin but in general, there may be many---despite
the fact that the lexmin is unique. We conclude the preliminaries by
relating isomorphism to lexicographic representatives.

\begin{observation}\label{observation:iso}
  Magmas $A=(D,\SRC)$ and $B=(D,\DST)$ are isomorphic iff their
  lexicographically smallest representatives are equal.
\end{observation}


The isomorphism problem for finite magmas is
graph-isomorphism-hard (GI-hard) even if we consider only
semigroups~\cite{zemlachenko82}. Further, deciding whether an incidence matrix
of a graph is lexmin is NP-hard~\cite{babai83}, despite the fact that GI is believed to be
easier than NP\@. Therefore, we do not expect the \lexmin problem for general magmas to
be computationally easy.

\section{Explicit Encoding}\label{sec:explicit}

A straightforward approach to the lexmin problem is to encode to SAT that a
target (unknown) magma $(D,\DST)$  is isomorphic to a given
magma $(D,\SRC)$.
Then, we can apply standard algorithms for finding the lexicographically
smallest magma $(D,\diamond)$, cf.~\cite{nadel-tacas16,phd/basesearch/Trentin19,petkovska-iccad16,silva-amai11}.

Effectively, $(D,\diamond)$ is represented in 1-hot encoding. First represent
an isomorphism $f\,:\,D\to D$ by introducing Boolean variables $\xpi{i}{j}$
meaning that $f(i)=j$ for $i,j\in D$.  Second, introduce additional Boolean
variables $\xn{i}{j}{v}$ meaning that $\dst{i}{j}=v$.


To ensure that the $\xpi{i}{j}$ variables represent a
bijection, generate cardinality constraints (converted to CNF
by standard means~\cite{handbook:cardinality}).

\begin{equation}\label{eq:bijection}
  \begin{split}
    \bij&(D):=\\
         & \left\{\sum\nolimits_{j\in D} \xpi{j}{i}=\sum\nolimits_{j\in D} \xpi{i}{j}=1\mid i\in D\right\}
  \end{split}
\end{equation}

To ensure that $\xn{i}{j}{v}$ represent an isomorphic $(D,\DST)$,
generate implications covering possible mappings between rows, columns, and
values.

\begin{equation}\label{eq:image}
  \begin{array}{l}
  \left(\xpi{r}{r'}\land\xpi{c}{c'}\land\xpi{\src{r}{c}}{v'}\right)\Rightarrow \xn{r'}{c'}{v'},\\
  \text{ for } r,r',c,c',v'\in D
  \end{array}
\end{equation}

\noindent Note that for row $r$ and column $c$, the value $\src{r}{c}$ is given.

An advantage is that we can easily apply any bitlevel lexicographic
optimization algorithms over the vector of variables representing the magma $(D,\DST)$, in the following order
$\xn{1}{1}{n},
 \xn{1}{1}{n-1},
 \dots,
 \xn{1}{1}{1},
 \xn{1}{2}{n},
 \dots,
 \xn{n}{n}{1}$.
A significant disadvantage is the sheer size of the encoding, which  involves
$\Theta(|D|^5)$ clauses. Therefore we propose a solution where the explicit
representation of $(D,\DST)$ is not necessary.


\section{Gradual Construction}\label{sec:gradual}
Instead of introducing variables for the unknown $(D,\DST)$,
we construct it gradually starting from its top-left corner,
continuing by filling the first row and then the second, and so on.
Here we avail of the concept of isomorphic copy, which is a magma induced by an isomorphism.

\begin{definition}[isomorphic copy]\label{def:copy}
   Consider a magma $(D_1,\SRC)$ and a bijection $f:D_1\rightarrow D_2$
   then the \emph{isomorphic copy} $(D_2,\DST)_f$ is defined as
   $\dst{a}{b}=f(\src{f^{-1}(a)}{f^{-1}(b)})$.
   In the remainder of the paper, we omit the subscript $f$ from $(D_2,\DST)_f$, whenever it is clear from the context that $f$ is present.
\end{definition}

The intuition behind an isomorphic copy is that to obtain the value $\dst{a}{b}$,
we first obtain the pre-images of $a$ and $b$,
then apply the (known) operation $\SRC$ to the pre-images in the context of $(D_1,\SRC)$,
and finally map the result back to~$(D_2,\DST)$. This is well-defined because $f$ is a bijection.

\begin{observation}
  Magmas $A=(D_1,\SRC)$ and $B=(D_2,\DST)$ are isomorphic iff there exists a
  bijection $f:D_1\rightarrow D_2$ such that $B$ is an isomorphic copy of $A$ by $f:D_1\rightarrow D_2$.
\end{observation}

To construct $(D_2,\DST)$, we will need to encode the constraints of the shape $\dst{r}{c}=v$, e.g.,
$\dst{1}{1}=1$ means placing 1 in the top left corner of the multiplication
table. Since $(D,\DST)$ must be an isomorphic copy of~$(D,\SRC)$, 
the constraint $\dst{r}{c}=v$ can be written as follows:
\begin{equation}\label{eq:pi} f(\src{f^{-1}(r)}{f^{-1}(c)})=v, \end{equation}
where $f$ is an unknown permutation of $D$.
As in the previous encoding, we encode $f$
as Boolean variables $\xpi{i}{j}$ coupled with the appropriate
cardinality constraints (see~\eqref{eq:bijection}).
The equality~\eqref{eq:pi} yields a set of implications covering all possible
values of $f$.\footnote{The implementation avoids repeated and tautologous clauses.}
\begin{equation}\label{eq:copyBool}
  \begin{split}
    \encode&(\dst{r}{c}=v) :=\\
           &\{\left(\xpi{i}{r}\land\xpi{j}{c}\right)\Rightarrow\xpi{\src{i}{j}}{v} \mid i,j\in D\}
  \end{split}
\end{equation}

Algorithm~\ref{algorithm:basic} shows how the lexmin
representative is calculated by maintaining a set of equalities $A$ of the
form $\dst{r}{c}=v$ for which we already know that they must hold in the
multiplication table of~$(D,\DST)$ (this is a loop invariant of the outer
loop). The inner loop attempts to extend the set of assignments $A$ for the next
cell of the multiplication table going from 1 to higher values. The call to the
function \encode conjoins the encoding of the assignments according to the
equation~\eqref{eq:copyBool} along with the bijection constraints~\eqref{eq:bijection}.

The algorithm first tries placing~$1$ in the top left corner and if that is possible it moves
onto the next column. Otherwise, it tries placing 2 in the top left
corner, and so forth. Once it succeeds in placing a value in a cell, the
value is fixed. The
algorithm leads to~$O(|D|^3)$ SAT calls. The permutation $f$, represented by the
Boolean variables $\xpi{i}{j}$, spans all permutations and therefore enables the
creation of any isomorphic copy of~$(D,\SRC)$ on the domain $D$. This also
justifies termination of the inner loop because one of the SAT calls is bound
to succeed since the set of isomorphic copies is always nonempty---it, for
instance, contains the input magma itself. Since $|A|\in O(|D|^2)$ and~\eqref{eq:copyBool}
requires $O(|D|^2)$ clauses, Algorithm~\ref{algorithm:basic} requires space for $O(|D|^4)$ clauses. 


\begin{algorithm}[t]
  \caption{Calculate lexmin $(D,\DST)$ for given $(D,\SRC)$
  by gradual construction.}\label{algorithm:basic}
  $A\gets \emptyset$ \tcp*[r]{empty set of assignments}
  \For{$r,c\in 1..|D|,1..|D|$}{
      $v\gets 1$\;
      \While{$\lnot\SAT(\bij(D)\cup\encode(A\cup\{\dst{r}{c}=v\}))$}{
        $v\gets v + 1$
      }
      $A\gets A\cup\{\dst{r}{c}=v\}$\tcp*[r]{extend $A$}
    }
\end{algorithm}

\section{Efficiency Improvements}\label{sec:improvements}

Algorithm~\ref{algorithm:basic} faces two major pitfalls: a high number of
SAT calls, and hard individual SAT calls. The upper bound of $O(|D|^3)$ on SAT
calls in Algorithm~\ref{algorithm:basic} is tight. For instance, for
quasigroups (aka Latin squares) it is also $\Omega(|D|^3)$.\footnote{Each row
of a quasigroup contains all the elements of $D$, therefore each row requires
$\frac{n(n-1)}{2}$ SAT calls as $v$ calls are needed for a cell containing the value $v$.} The second issue, where an individual SAT call
might be too hard, is potentially even more worrisome.

Indeed, we have a reason to believe that some SAT calls will be hard due to an
underlying pigeonhole principle. For instance, if the original magma $(D,\SRC)$
does not contain any element more than $k$-times on any given row, the same
must hold for the target magma $(D,\DST)$. Then, for the SAT solver to prove
that it cannot place an element for the $k+1$-th time on the same row is indeed
reminiscent of the pigeonhole principle formulas, which are well known to be
difficult for SAT (and resolution in general)~\cite{Haken85,RezendeNR020}.
Such hard SAT calls could get the Algorithm~\ref{algorithm:basic} simply stuck on a single cell.

Here we focus on designing new propagation techniques that let us bypass calls
to the SAT solver in specific scenarios. We focus mainly on techniques that
rely on counting because that is a famous Achilles' heel for modern SAT
solvers. We begin with a technique that enables in some cases identifying the first row.

\subsection{Identification of the First Row}\label{sec:first}

Recall that any row $r$ of the original magma $(D,\SRC)$ must be projected to
some row $r'=f(r)$ in the target magma.  Here we show that in certain cases  it
is possible to identify possible candidates that might be mapped to the
first row, i.e., we construct a set $C_1\subseteq D$, s.t.\, $f(a)=1$ only if $a\in
C_1$. This is encoded into the SAT solver as a set of unit clauses:
$\{\{\lnot\xpi{a}{1}\}\mid a\notin C_1\}$ before
Algorithm~\ref{algorithm:basic} starts.

Suppose that $\src{4}{x}=4$, for all $x\in D$, i.e., the $4^\text{th}$ row is
entirely filled with $4$'s. If $4$ is renamed to $1$, i.e., pick an isomorphic
copy with $f(4)=1$, the first row of $\DST$ becomes all 1's, i.e.,
lexicographically smallest first row possible. We generalize this idea to find
candidates for the first row of~$\DST$.

\begin{definition}
Let $A=(D,\SRC)$ be a magma with some idempotents. The \emph{idempotent apex} of
$A$ is the largest value of $|\{x\in D \mid \src{e}{x}=e\}|$, for $e\in D$ idempotent
of~$A$.
\end{definition}

Possible rows that can be mapped to the first row are obtained by calculating
for each row $r$ of $\SRC$ that contains an idempotent, how many times $r$
appears in it, i.e., $o_r := |\{c\in D \mid \src{r}{c}=r\}|$, if
$\src{r}{r}=r$. We claim that only a row that maximizes this number can become
the first row in the smallest representative $\DST$, i.e., $f(r)=1$ implies
$o_r$ is the apex of the input magma. If the input magma does not contain any
idempotents, this technique is not applied. Note that in the example of
Figure~\ref{fig:example} only row 4 contains an idempotent and therefore
it necessarily must become the first one.

We proceed with the correctness proof of this statement. For succinctness
we introduce the following notation. We write $[(D,*)]$ for the set of
isomorphic copies $(D,\DST)$ isomorphic to $(D,*)$. We write $\down (D,*)$
for the lexicographically smallest representative according to the ordering
$\prec_r$ (by-rows). We write $1\DST\{1,\ldots,k\}=\{1\}$ as a shorthand
for $1\DST i=i$, for $i\in 1..k$, which effectively means that the first $k$
columns of the first row of $\DST$ are equal to 1.

\begin{proposition}
  Let $A=(D,*)$ be a magma with idempotents and idempotent apex $k$. Let 
  $\mathcal{M}_k:=\{(M,\DST)\in [(D,*)]\mid 1\DST\{1,\ldots,k\}=\{1\}\}$.
  Then 
  \begin{enumerate}
    \item $\mathcal{M}_k\neq \emptyset$; 
    \item $\down (D,*)\in \mathcal{M}_k$. 
  \end{enumerate}
\end{proposition}
\begin{proof}
  Let $e\in D$ such that $\src{e}{e}=e$ and $D_0:=\{x\in D \mid e * x = e\}$ has size $k$. Pick
  $g$, a permutation of $D$, such that $g(D_0)=\{1,\ldots,k\}$ and $g(e)=1$.
  Define on $D$ the following operation: $\dst{x}{y}:=g(g^{-1}(x) * g^{-1}(y))$,
  for all $x,y\in D$. For all $x\in \{1,\ldots,k\}$, we have
  \[
    \dst{1}{x} = g(g^{-1}(1) * g^{-1}(x))=g(e*g^{-1}(x))=g(e), 
  \]
  because $g^{-1}(x)\in D_0$ and $e*a=e$, for all $a\in D_0$. It is proved that
  $\dst{1}{x}=g(e)=1$, for all $x\in\{1,\ldots,k\}$. In addition,
  $\dst{x}{y}:=g(g^{-1}(x) * g^{-1}(y))$ implies that (replacing $x$ with $g(x)$
  and $y$ with $g(y)$) $\dst{g(x)}{g(y)}=g(g^{-1}(g(x)) * g^{-1}(g(y)))=g(x * y)$. It
  is proved that $g$ is an isomorphism of the magmas $(D,\DST)$ and $(D,*)$.
  Therefore $(D,\DST)\in \mathcal{M}_k$. The first claim follows. 

  
  Regarding the second claim, suppose that $(D,\times)$ is a lexmin of $(D,*)$.
  Since $M_k$ is not empty, we must have $1\times j = 1$ for all $j$ in
  ${1,\dots,i}$ and some $i\ge k$. Since the idempotent apex is preserved by
  isomorphism (see Observation~\ref{observation:invariant}), we have $i\le k$.
  Hence $i=k$ and $(D,\times)$ is in~$M_k$.
\end{proof}

\subsection{Budgeting}\label{sec:budgeting}
Next, we describe a technique  that is invoked for every SAT call of
Algorithm~\ref{algorithm:basic}. Roughly speaking, each element $a\in D$ is assigned a
budget, which is decremented whenever $a$ is placed in the target table.  SAT
calls $\dst{r}{c}=v$ with values $v$ that have 0 budget are not invoked (and
deemed unsatisfiable). We consider budgets per row/column or for the whole table.
In the context of constraint programming, similar propagation
techniques are abundantly used for global constraints~\cite[Chapter~3]{cp-handbook}.

For intuition, consider a situation where each row of the multiplication table of $\SRC$
contains \emph{at most one} occurrence of any given element (as in the example Figure~\ref{fig:example}). Then the same property
will hold in the rows of $\DST$ by Observation~\ref{observation:invariant}. This means that if
Algorithm~\ref{algorithm:basic} has placed an element~$a$ in a certain row, it
does not need to try placing it in the same row again. This enables the
algorithm to skip SAT calls on values that are no longer possible (in that row).

This idea is readily generalized to an arbitrary number of occurrences. Define
$o_{\SRC}(r, a)=\left|\{c \mid \src{r}{c}=a,\,c\in D\}\right|$ and calculate
$\max\{o_{\SRC}(r,a)\mid r,a\in D\}$ to give a budget for an arbitrary element in
an arbitrary row of $(D,\DST)$. The same can be applied to columns and the
total number of occurrences in the table. This is especially useful for
quasigroups, where each element appears precisely once in each row/column.

The budget calculated as described above is an upper bound, which can sometimes
be improved. Consider the case when the first row was uniquely identified by the
technique outlined in the previous section. Then we have established that
$f(k)=1$ for some $k\in D$, for any $f$ yielding the lexmin copy.
This enables splitting budgets for the element 1 and the
rest of the elements according to the following equalities.
\begin{align}
  \begin{split}
    \max\{o_{\SRC}(r, k)\mid r\in D\} = \max\{o_{\DST}(r, 1)\mid r\in D\}
  \end{split}\\
  \begin{split}
    \max\{o_{\SRC}(r, a)\mid a\neq k\land r,a\in D\} ={}&\\
    \max\{o_{\DST}(r, 1)\mid a\neq 1\land r,a\in D\}
  \end{split}
\end{align}

\subsection{Row Invariants}\label{sec:invariants}
\newcommand{\invariants}{\mathcal{R}} 
\newcommand{\invariant}{I}

As shown above, the budgeting technique can benefit from knowing which element
has been mapped to the first row. More generally, once it is established that
$f(k)=j$,  for some $k,j\in D$, it must hold that the number of occurrences of $k$ in $(D,\SRC)$ will
be equal to the number of occurrences of $j$ in the copy~$(D,\DST)$. But how to
establish such correspondence? Note that the variables $\xpi{i}{j}$ determine
the permutation on the elements of $D$ but this permutation may change over
the course of the algorithm.


From the definition of isomorphic copy (Definition~\ref{def:copy}), the contents of a row of the
original table of $(D,\SRC)$ must correspond to the contents of some row of the
table of $(D,\DST)$. More precisely, the bag of elements $[\src{r}{c}\mid c\in D]$
is equal to the bag of elements $[\dst{f(r)}{c}\mid c\in D]$. In some cases,
this lets us unequivocally identify that a row $r$ in the original magma maps to a
row $r'$ in the isomorphic copy. This is done by calculating \emph{invariants}
(properties invariant under isomorphism) and matching pairs of rows with unique
invariants. Currently, we use the following invariants bundled into a
single one. Similar invariants have been used before for isomorphism
testing~\cite{cp21,constraints22boosting,loops3.4.1}.

\begin{itemize}
  \item $|\{r\circ c = c\mid c\in D\}|$, for fixed $r\in D$ and $\circ\in\{\SRC,\DST\}$
  \item $|\{r\circ c = r\mid c\in D\}|$, for fixed $r\in D$ and $\circ\in\{\SRC,\DST\}$
  \item $|\{r\circ r = r\}|$, for fixed $r\in D$ and $\circ\in\{\SRC,\DST\}$
  \item define $g_r(a)=r\circ a$ and
    $m_r(a)$ as minimal $k$ s.t.\ $g_{\SRC}^k(a)=g_{\SRC}^j(a)$ for some $j<k$.
    Take the bag $[m_r(c)\mid c\in D]$ as invariant, for fixed $r$ and $\circ\in\{\SRC,\DST\}$.
\end{itemize}
For the example in Figure~\ref{fig:example}, only row 4 has 7 columns $c$ s.t.\ $\src{4}{c}=c$
and $m_4(c)=1$.
The invariants are used in Algorithm~\ref{algorithm:basic} as follows. Each time
a row $r$ of $(D, \DST$) is entirely filled, its invariant is calculated and if there is
a \emph{unique} row $r'$ in the input table $(D, \SRC)$ with the same invariant, set
$f(r')=r$, add the corresponding unit clause $\{\xpi{r'}{r}\}$ and recalculate budgets.

We also exploit invariants even if they do not give us a unique correspondence
of rows. In the case that an invariant is shared by $k$ rows in $\SRC$ and it
already appears $k$ times in the partially filled copy~$\DST$, subsequent rows
will never be mapped to the ones that gave rise to the invariant in question.
More concretely, if there is a set of rows $R\subseteq D$ with $|R|=k$ that
correspond to a certain invariant $\invariant$ and the invariant $\invariant$
already appears $k$ times in the first $r$ rows of $\DST$ then for $f(r')\neq
j$ for $j\in R$ and $r'>r$. In the implementation, corresponding unit clauses
are inserted into the SAT encoding once that takes place. We remark the same
technique could be applied to columns but it would not be useful since columns
are never complete until the very end.

\subsection{Mid-Row Budgeting Refinement}\label{sec:budgeting:idem}
The techniques described in the previous section enable refining budgets after
a row of the target table is filled. Here we also show that this can be done
mid-row.
We propose a cheap technique that is easy to implement where we split the rows
of $\SRC$ into rows containing an idempotent and into rows that do not. Note
that row $r$ contains an idempotent iff $\src{r}{r}=r$. This lets us calculate
three types of budgets: (1)~for all rows of $\SRC$; (2)~for rows of $\SRC$
containing an idempotent; (3)~for rows of $\SRC$ not containing an idempotent.

When Algorithm~\ref{algorithm:basic} starts filling a row $r$,
it does not know in which group the row falls and therefore starts with the global
budget. Once the $r^{\textrm{th}}$ position is filled, the budget can be
refined accordingly. In the row-based traversal, in the first row, the
refinement happens once the top left corner has been filled (the first column of
the first row).

\subsection{Upper Bound by Last Value}\label{sec:last}
A simple improvement is obtained by inspecting the model obtained from
satisfiable SAT calls. 
Even though Algorithm~\ref{algorithm:basic} only imposes
assignments to the table $\DST$ for those cells that have been traversed so far,
any SAT model represents a permutation for all the elements in the domain $D$,
from which one can infer the rest of the table of~$\DST$. The remainder (untraversed) of
the table does not necessarily guarantee that it is lexicographically smallest
but it gives us an upper bound. This means that for each cell $(r,c)\in D\times
D$ there is always a tentative value $v_{u}$ for which we already have a
witnessing permutation. This lets us avoid the SAT call for the query
$\dst{r}{c}=v$ for any $v\geq v_{u}$. This upper bound is also used in different
search strategies described in the upcoming section.
We remark that an analogous technique has also been used for explicit representation-based
calculation of lexicographically smallest SAT assignment~\cite{knuth-sat}.

\subsection{Search Strategies}\label{sec:strategies}

Algorithm~\ref{algorithm:basic} performs $|D|$ tests for a single cell of the
table $\DST$ in the worst case. It is tempting to apply standard techniques
for minimization, such as binary search. However, these are not directly
applicable because the behavior is not monotone, e.g., it might be possible
to place~3 and~7 in a specific cell, but not 5. Nevertheless, monotone behavior
can be obtained by constructing SAT queries over a \emph{disjunction} of values.
Hence, instead of querying $\dst{r}{c}=v$, we query $\bigvee_{v\in
V}\dst{r}{c}=v$ over some set of $V\subseteq D$. In terms of the SAT encoding, one
could calculate a disjunction over the encoding for a single value
(equation~\eqref{eq:copyBool}) but we are able to avail of the common part and $\dst{r}{c}\in V$
is encoded as follows.
\begin{equation}
  \{ \left(\xpi{i}{r}\land\xpi{j}{c}\right)\Rightarrow\bigvee\nolimits_{v\in V}\xpi{\src{i}{j}}{v}, \mid i,j\in D\}
\end{equation}

This approach has monotone behavior in the sense that if
$\dst{r}{c}\in V$ is satisfiable then also $\dst{r}{c}\in V'$ is satisfied for
any $V\subseteq V'$. This enables us to use standard MaxSAT iterative techniques,
where the basic Algorithm~\ref{algorithm:basic} is in fact a linear UNSAT-SAT
strategy. Additionally, taking into account values obtained from satisfiability
calls enables improving the upper bound for linear
SAT-UNSAT or binary search.

In our experiments, standard binary search did not perform well because it still
requires $\Omega(\log_2 |D|)$ SAT calls to prove an optimum. Therefore we apply
a modified binary search where first we test if the optimum has not already been
reached. In the case that the optimum has not been reached, the upper bound is
updated. If the upper bound reduced the search space by a factor of 2, we simply
recur. If the upper bound falls into the top half of the possible values,
another SAT call is issued.

\section{Experiments}\label{sec:experiments}

The experiments are run on an Intel\textregistered\ Xeon\textregistered CPU
E5-2630 v2 \@ 2.6 GHz $\times$24 computer, with 64 Gb RAM\@. We call our tool
\mlex and it supports two SAT solvers, \texttt{minisat}~\cite{minisat} and
\texttt{cadical}~\cite{cadical}. Unless otherwise stated, \texttt{minisat} is
used in our experiments. Both SAT solvers are used incrementally and
\texttt{cadical} is used via the \texttt{IPASIR}
interface~\cite{IPASIR}. We excluded the Explicit Encoding from the evaluation
since it led to unwieldy memory consumption (dozens of gigabytes even
for small problems). The GAP package Smallsemi~\cite{smallsemi}
provides a function to calculate
lexmin semigroup, which is not included in the evaluation
because the ordering used traverses the table by the diagonal first and the
implementation suffers from timeouts and large memory consumption
even on small problem instances (order 20).
Hence, the evaluation is based on
Algorithm~\ref{algorithm:basic} and its extensions described in the Efficiency
Improvements section.

The tool was evaluated on
several popular algebraic structures
(algebras) defined in Table~\ref{table:algebra:mace4:definition}.
In this
table, ``$e$'' is a constant, ``$*$'' is a binary operation, and ``$'$''
unary; all clauses are implicitly universally
quantified. Even though \mlex currently supports only a single binary
operation, it can handle all these algebras. This is because in many
finite algebraic structures, such as those listed here,
the constant and the unary function
are uniquely determined by the binary operation. Hence,
 they can be removed from the inputs to \mlex.

\begin{table}[tb]
\begin{tabular}{p{1.8cm}p{5.6cm}}
	\toprule
	Structure  & Definition in FOL   \\\midrule
	Groups & $x*(y*z)=(x*y)*z$,\ \ $x * e = x$,\ \ $x * e = x$,\ \ $x * x' = e$,\ \ $x' * x = e$\\
	Loops & $x*y=x*z\rightarrow y=z$,\ \ $y*x=z*x \rightarrow y=z$, $e*x=x$, $x*e=x$\\
	Quasigroups & $x*y=x*z\rightarrow y=z$,\ \ $y*x=z*x\rightarrow y=z$ \\
	Semigroups & $x*(y*z)=(x*y)*z$ \\
	Magmas & no requirement \\ \bottomrule
\end{tabular}
\caption{FOL definitions of the used algebraic structures.}%
\label{table:algebra:mace4:definition}
\end{table}

The evaluation was performed on randomly generated samples from five algebraic structures: groups, loops, general magmas,
quasigroups, and semigroups.
For groups, we randomly pick
the groups given by the \texttt{AllSmallGroups} function in GAP\@. For magmas
and semigroups, we generate them with the help of GAP functions such as
\texttt{Random}.  For quasigroups and loops, we use the
\texttt{RandomQuasigroup} and \texttt{RandomLoop} functions in the
LOOPS package in GAP\@. We make sure the models in each structure do not
belong to a sub-structure in the list above.  For example, the magmas we use are
not semigroups or quasigroups. We consider a total of 210 random samples of the
five algebraic structures listed in Table~\ref{table:algebra:mace4:definition}, of orders 16
to 128 in increments of 16. In addition, we include random samples of 5 magmas of each of the orders 192 and 256.
Finally, a timeout of 30 minutes is used for calculating the lexmin copy of each model.




\subsection{Ablation Study of Techniques}

We test the introduced techniques in an ablation study. We
consider basic Algorithm~\ref{algorithm:basic}, a version with all
improvements turned on, and the effect of turning off each one of them
individually. For search strategies, we compare between
linear-unsat-sat (\texttt{lus}) and modified binary search (\texttt{bin2}).

\begin{figure}[tb]
  \centering
	\includegraphics[width=.99\columnwidth]{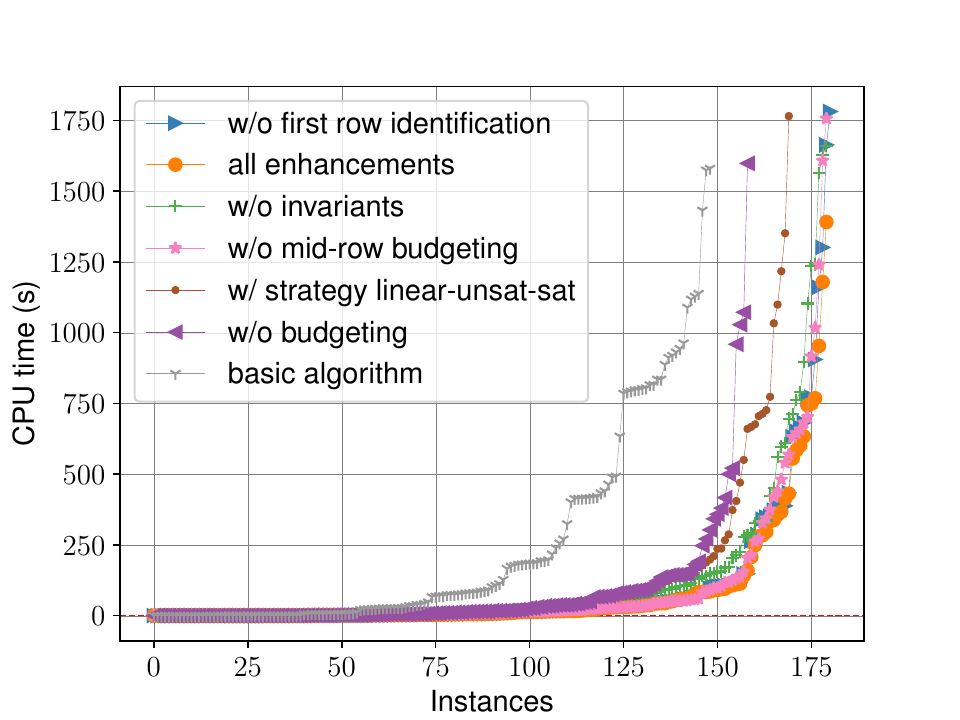}
	\caption{Performance of \texttt{mlex} with different options.}%
	\label{fig:algebras:mlex:time}
\end{figure}

Figure~\ref{fig:algebras:mlex:time} shows a cactus plot for the ablation study.
Although all the techniques lead to an improvement in the
tool, the most significant is the use of budgeting, which confirms our
suspicion that hard SAT calls might occur due to counting arguments.
Interestingly, the binary search technique also has a significant impact.
Turning off  the other techniques does not have a significant impact on the
number of solved instances. However, there are specific classes of problems that cannot be
solved without using all the techniques. Also, the ``all enhancement'' version of
the solver appears to be the fastest and the most robust version.



\begin{figure}[!htb]
  \centering
	\includegraphics[width=0.99\columnwidth]{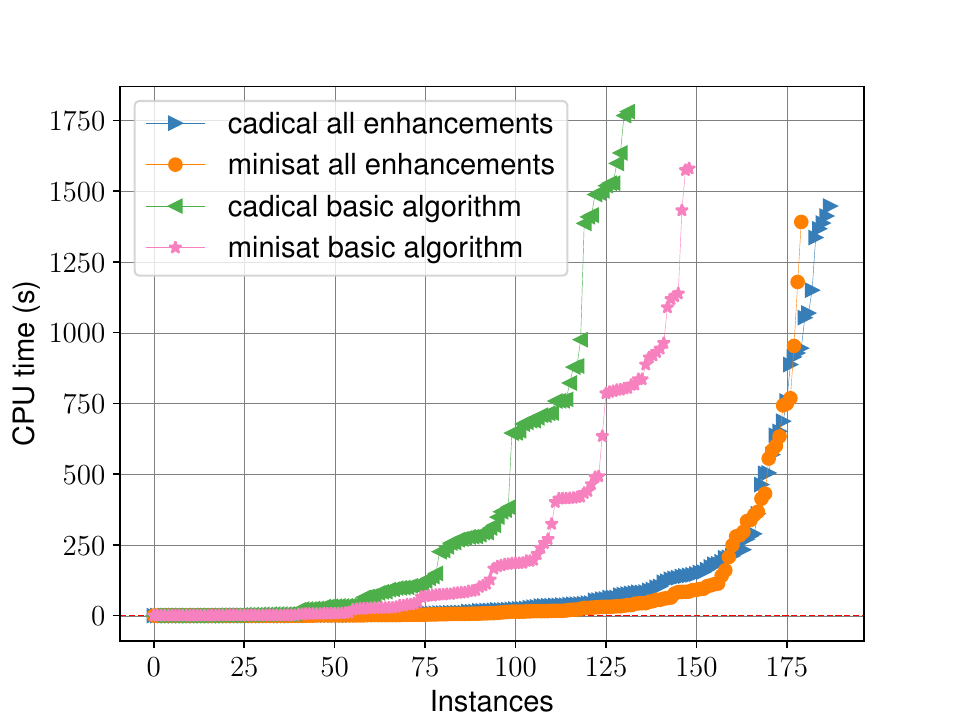}
	\caption{Comparison of \texttt{minisat} and \texttt{cadical} in \mlex.}%
	\label{fig:algebras:mix:sat:solvers:time}
\end{figure}

It is well-known that \texttt{minisat} is simple and fast and that for more
complex problems, \texttt{cadical} usually performs much
better~\cite{dutertre2020empirical}. This pattern is also observed with \mlex.
As shown in the cactus diagram Figure~\ref{fig:algebras:mix:sat:solvers:time},
when enhancement features are turned on, then for simpler problems that take a
shorter time, \texttt{minisat} usually solves more problems for the same time,
but for more complex problems, the opposite is true.  However, as also shown in
the same diagram, the choice of other input options to \mlex has a much more
pronounced impact on the speed of \mlex than the underlying SAT solver, as the
curves corresponding to both SAT-solvers are very close for the same set of
input options. Surprisingly, \texttt{cadical} performs poorly compared to \texttt{minisat} when all improvements are turned off.

\section{Related Work}\label{sec:related}


Finite model finding is ubiquitous to automated reasoning.
Sometimes, users are interested in models rather than in proving a
theorem~\cite{mccune94}. In theorem proving, models serve as
counterexamples to invalid conjectures~\cite{blanchette-lpar10}, which also
appear in software verification~\cite{torlak-tacas07}. Finite models have also been used as a semantic
feature for \emph{lemma selection learning}~\cite{urban-ijcar08}.
In certain fragments, finite model finding provides a complete decision
procedure, e.g., the \emph{Bernays-Sch{\"o}nfinkel} fragment (EPR). Throughout
the years, CP, SAT, and SMT tools have been used in finite model
finders~\cite{AudemardBH06,claessen03,reynolds-cav13,reynolds-cade13,ZhangJ1996Falcon,ZhangZ95,cubes}.
SAT and CP are routinely used to solve algebraic problems~\cite{heule-aaai18,semigroups10,generators}.


It is important to note that finite models are also constructed by dedicated
approaches based on deep domain knowledge. Notably, the algebraic system
GAP~\cite{GAP4} contains a number of packages for specific types of algebraic
structures. The Small Groups library~\cite{besche-ijac02} contains
\emph{all} ($\approx 4\times 10^8$) non-isomorphic groups up to order 2000 (except for order 1024).
Similarly, Smallsemi~\cite{smallsemi} catalogues semigroups and LOOPS
packages loops~\cite{loops3.4.1}. However, currently, these packages do not
provide the lexicographically smallest representative. Adding our tool into
GAP is a subject of future work.

Normal forms are ubiquitous in computer science and mathematics. Here we highlight
the canonical labeling algorithms implemented in the
\nauty system~\cite{nauty}. The system has been developed since the 80's
and it is considered state-of-the-art for graph isomorphism (and more).
It is possible to construct a canonical form of a magma by using \nauty:
for a magma $A$, construct a special graph $G'_A$ and find its canonical graph $G_A$, cf.~\cite{khanIEEE20}.
This form is canonical in the sense that two isomorphic magmas will give the
same canonical graph but the resulting graph is opaque to the user. Hence, it cannot be used for solving the problem tackled in this paper.

A large body of research exists on \emph{symmetry breaking} in SAT and
CP~\cite{cp-handbook,sakallah21}. In general, however, the objective of
symmetry breaking is different from our objective: it is a means speeding up
search by avoiding symmetric parts of the search space. In contrast, in our
case, the normal form is the objective. Typically, symmetry breaking is meant to
be fast, when used dynamically, or should add a small number of constraints,
when used statically~\cite{Codish2018,codish-aaai20}. Therefore, symmetry
breaking is often incomplete. Even though, Heule investigates optimal complete
symmetry breaking for small graphs~($\approx 5$ vertices)~\cite{heule-mcs19}.
\citet{szeider_cp21} develop a specific symmetry breaking,
called~\emph{SAT Modulo Symmetries}, where a SAT solver is enhanced to
look for the lexicographically smallest graph (similarly to
lazy SMT). There, the objective is to enumerate non-isomorphic graphs
with certain properties.
More broadly, this paper fits into the SAT+CAS paradigm, where SAT is
combined with \emph{computer algebra systems}, cf.~\citet{bright-cacm22}.
\section{Conclusions and Future Work}\label{sec:conclusions}
This paper tackles the problem of calculating the lexicographically smallest
representative of a given algebraic structure. This is a fundamental problem in
computational algebra, where the user, a mathematician, requires a \emph{specific
canonical form}.  A prominent feature of this canonical form is that it enables
a ``common language''  between different mathematical libraries and it
enables the mathematicians to identify familiar patterns and structures.

Our prototype of the proposed algorithms shows that the SAT technology is up
to the task. The proposed encoding enables tackling large problem instances by
avoiding explicitly representing the target structure. The SAT solver is used
in a black box fashion with repeated SAT calls, which gradually construct the
targeted structure (the lexicographically minimal representative). We further
design a number of dedicated techniques that enable simplifying, or completely
avoiding, certain SAT calls.  The experimental evaluation shows that the
approach decidedly benefits from  this additional propagation (done outside of
the SAT solver).


This work opens a number of avenues for further research.
More powerful propagation techniques still could be considered---such as different invariants
and more aggressive and fine-grained propagation.
A tighter integration with the SAT solver and application to structures with
several multiplication tables is more of an engineering effort but would
further increase the practicality of the implemented tool. Rather than invoking
the approach on a \emph{given} structure, it would also be interesting to
integrate it into the calculation of non-isomorphic structures under
constraints.

\section*{Acknowledgements}
%
We thank Brendan D.\ McKay for helpful comments.
The results were supported by the Ministry of Education, Youth and Sports within the dedicated program ERC~CZ under the project \emph{POSTMAN} no.~LL1902
and by national funds through the FCT --- Funda\c{c}\~{a}o para a Ci\^{e}ncia e a Tecnologia, I.P.,
under the scope of the project UIDB/00297/2020 (doi.org/10.54499/UIDB/00297/2020) and the project UIDP/00297/2020 
(doi.org/10.54499/UIDP/00297/2020) (Center for Mathematics and Applications) and
co-funded by the European Union under the project \emph{ROBOPROX} (reg.~no.~CZ.02.01.01/00/22\_008/0004590).
This article is part of the \emph{RICAIP} project that has received funding
from the European Union's Horizon~2020 research and innovation programme under
grant agreement No~857306.
P.~Vojt\v{e}chovsk\'y was supported by the Simons Foundation Mathematics and Physical Sciences Collaboration Grant for Mathematicians no.~855097.
%

\bibliography{refs,learning,fm,ja}
\end{document}